\documentclass[twocolumn,10pt]{article}

\usepackage[T1]{fontenc}
\usepackage{lmodern}
\usepackage[letterpaper,margin=0.6in,columnsep=22pt]{geometry}
\usepackage{amsmath}
\usepackage{booktabs}
\usepackage{graphicx}
\usepackage{microtype}
\usepackage{siunitx}
\usepackage{tabularx}
\usepackage{xcolor}
\usepackage{enumitem}
\usepackage[font=footnotesize,labelfont=bf]{caption}
\usepackage{titlesec}
\usepackage{balance}
\usepackage{tikz}
\usetikzlibrary{arrows.meta,positioning}

\usepackage[
  backend=biber,
  style=authoryear-comp,
  natbib=true,
  sorting=nyt,
  giveninits=true,
  maxcitenames=2,
  maxbibnames=99,
  useprefix=true,
  uniquename=false,
  doi=true,
  url=true,
  isbn=false
]{biblatex}
\usepackage{xurl}
\usepackage{xspace}
\usepackage[hidelinks]{hyperref}
\hypersetup{
  pdftitle={Gridnberg: A Topography-Aware Pedestrian Routing Dataset for New York City},
  pdfauthor={Ariel Noyman},
  pdfkeywords={pedestrian networks, elevation, slope, routing, accessibility, New York City}
}

\graphicspath{{figures/}}
\setlist{nosep,leftmargin=1.35em}
\titlespacing*{\section}{0pt}{0.9em}{0.3em}
\titlespacing*{\subsection}{0pt}{0.6em}{0.2em}
\titlespacing*{\paragraph}{0pt}{0.5em}{0.8em}
\AtBeginBibliography{\footnotesize}
\newcommand{\dataset}{\textsc{Gridnberg}\xspace}
\newcommand{\pct}{\,\%\xspace}

\title{Gridnberg: A Topography-Aware\\Pedestrian Routing Dataset for New York City}
\author{Ariel Noyman\\
  \small Massachusetts Institute of Technology, Cambridge, MA, USA\\
  \small \texttt{noyman@mit.edu}}
\date{}

\begin{document}
\twocolumn[
  \begin{@twocolumnfalse}
    \maketitle
    \begin{center}
      \begin{minipage}{0.78\linewidth}
        \begin{abstract}
          \noindent
          Cities are rarely flat, yet urban network analysis usually represents streets as planar graphs. This simplification affects modeled impedance, route choice, and the interpretation of accessibility, particularly where alternative paths differ in grade. This paper introduces \dataset ('grid-n-berg', grid and mountain), a topography-aware pedestrian routing dataset for New York City. The dataset enriches the NYCWalks network with vertex-level elevations derived from the New York City Planimetric Database. For each pedestrian-network geometry vertex, the workflow averages selected elevation observations within a \SI{50}{\meter} radius, retains segments with complete vertex support, and uses direction-specific cumulative ascent and descent to calculate three routing costs: horizontal distance, a comfort-oriented slope score, and an accessibility-sensitive slope score. The release retains \num{313184} of \num{315577} source segments (99.24\pct). \dataset supports reproducible terrain-aware analysis, transparent scenario comparison, and improved pedestrian-network representations in New York and other cities.
        \end{abstract}
        \vspace{0.35em}

      \end{minipage}
    \end{center}
    \vspace{0.8em}
  \end{@twocolumnfalse}
]

\section{Introduction}

\begin{figure*}[t]
  \centering
  \includegraphics[width=0.98\textwidth]{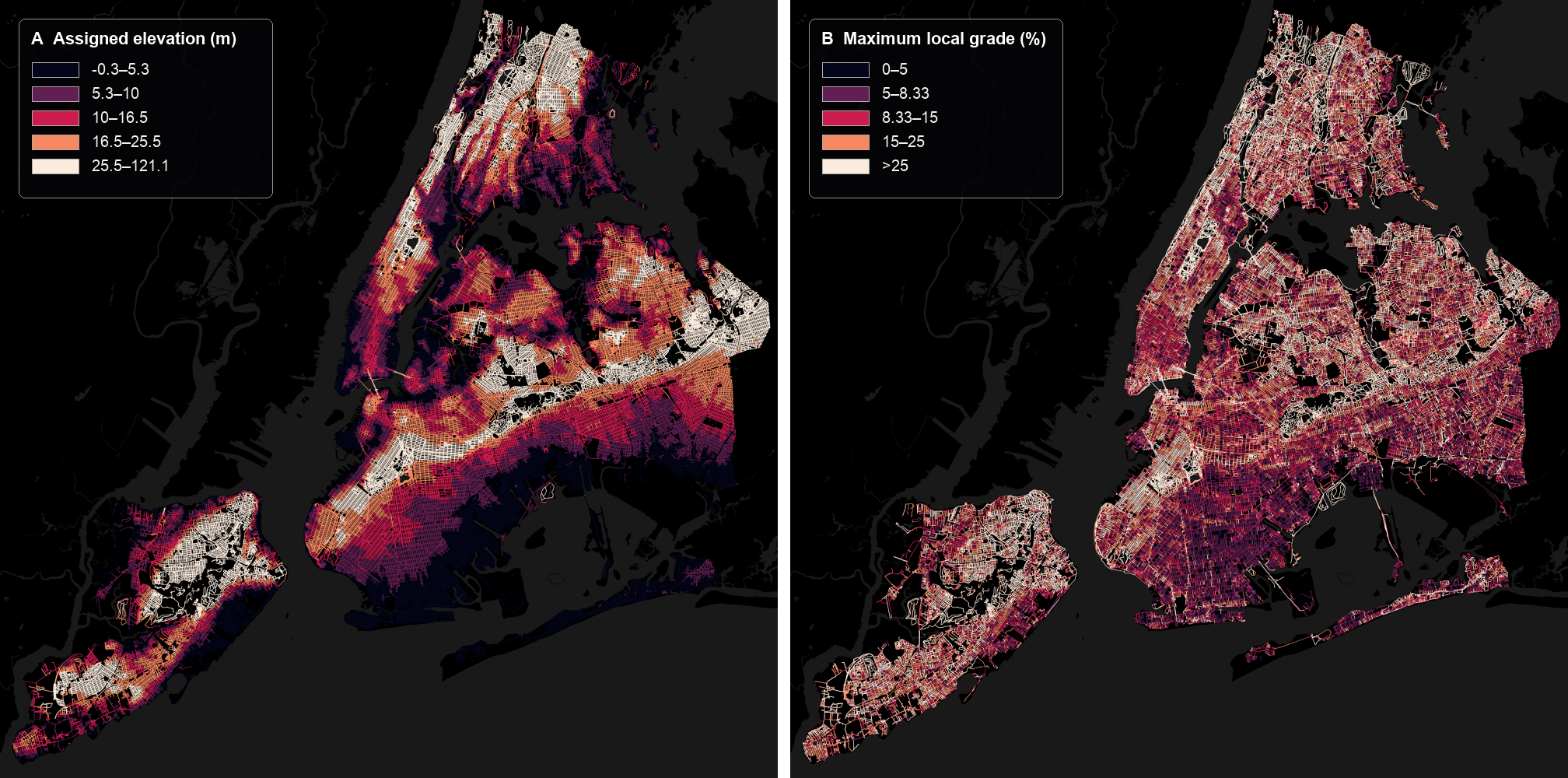}
  \caption{Comparative outputs of \dataset. Panel A maps assigned elevation at each stored segment's \(a\) endpoint (attribute \texttt{z\_a\_m}) spanning \SI{-0.3}{\meter} to \SI{121.1}{\meter}. Panel B maps the maximum reported absolute local grade, with breakpoints at 5, 8.33, 15, and 25\pct.}
  \label{fig:citywide}
\end{figure*}

At the city scale, many models begin by flattening the road network. Street segments become two-dimensional edges, route impedance becomes horizontal distance, and measures of centrality or accessibility are calculated on a plane. This abstraction is useful and has enabled reproducible, large-scale comparisons of urban form \citep{boeing2017osmnx,boeing2022street}. Nevertheless, walking unfolds over changes in elevation. Grade affects walking speed and metabolic effort and can alter the relative attractiveness of routes and modes \citep{minetti2002energy,kay2012route,meeder2017slope,campbell2019crowdsourced}. Even cities commonly conceived as flat contain hills, ramps, stairs, bridges, depressed roadways, and multilevel pedestrian connections.
\\
Sidewalk-specific networks improve spatial fidelity relative to road-centerline models \citep{rhoads2023sidewalk,hosseini2023mapping}. Unless enriched with elevation, however, a planar graph still treats links with equal horizontal length as equivalent even when one climbs sharply. Although some navigation services may incorporate elevation, their proprietary data and costs generally cannot be inspected or reproduced by researchers.
\\
Related work extends beyond planar road centerlines in two directions. Sidewalk-network research has improved the representation and validation of pedestrian infrastructure, including the citywide NYCWalks graph used here \citep{rhoads2023sidewalk,hosseini2023mapping,sevtsuk2026spatial}. Elevation-aware studies include profile and ``easiest path'' routing, three-dimensional pedestrian topology, incline-aware accessibility measures, and New York-specific sidewalk-slope analysis \citep{nourian2015easiest,rahaman2017capra,sun2021connecting,bolten2021routine,hosseini2024inclusive,franchi2025robotability}. \dataset brings these together in a public implementation with locally sampled municipal elevations, directional grade measures, and cost scenarios. Figure~\ref{fig:citywide} contrasts the citywide distribution of assigned elevation with the local-grade diagnostic reported for the released network.

\section{Data}

\subsection{Pedestrian Network Data}

The workflow combines two public sources with different geometries, coordinate reference systems, and semantic roles. The pedestrian source is the supplemental NYCWalks network created by \citet{sevtsuk2026spatial}. It contains \num{315577} LineString records representing sidewalks, crosswalks, and footpaths. The working copy is projected in EPSG:6538 (NAD83(2011) / New York Long Island) and uses metres. The source study also provides pedestrian-count estimates, but \dataset retains only the geometry.

\subsection{Elevation Data}

Elevation comes from the 2022 NYC Planimetric Database \texttt{ELEVATION} class \citep{nyc2022elevation,nyc2022capture}. The full class contains \num{1471855} point records whose horizontal coordinates use EPSG:2263 (NAD83 / New York Long Island in US survey feet); elevations are referenced to NAVD88 and are also recorded in US survey feet. Before matching, the workflow retains \num{387566} roadbed spot and bridge points identified by \texttt{FEATURE\_CODE=3000} and \texttt{SUB\_FEATURE\_CODE} values 300000 and 300020. Building-roof, water, and other feature classes are excluded to reduce contamination of the local mean.

The Planimetric capture rules place spot elevations at the beginnings, intermediate points, and ends of roadbed and interior-sidewalk-centerline features, generally at intervals no greater than 200 feet \citep{nyc2022capture}. They are therefore relevant local observations, but they do not measure cross slope, and pedestrian or bicycle bridges may lack dedicated bridge elevation points.

\section{Methods}

\subsection{Elevation Assignment}

\begin{figure}[t]
  \centering
  \includegraphics[width=\linewidth]{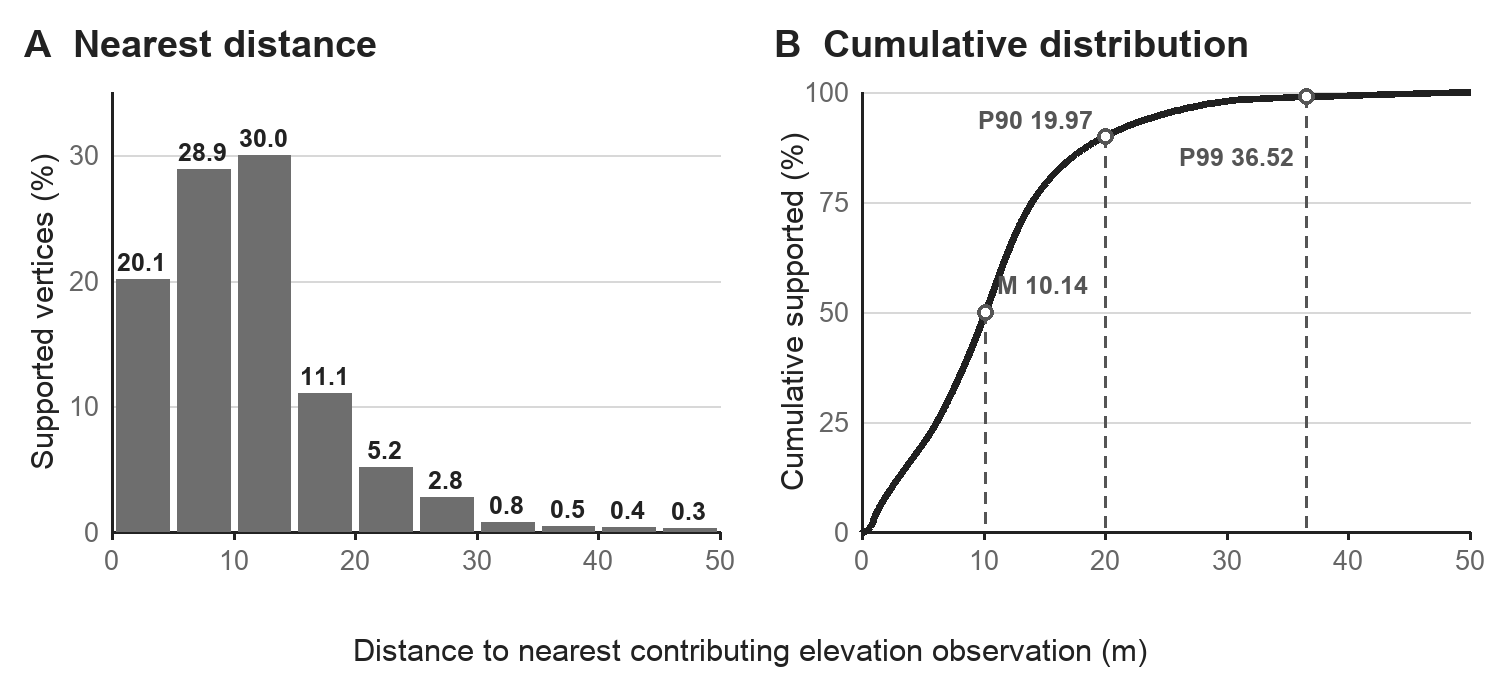}
  \caption{Nearest selected elevation observation for supported source-network vertices. Panel A shows \SI{5}{\meter} distance bins; Panel B shows the cumulative distribution and its median, 90th percentile, and 99th percentile.}
  \label{fig:support-distance}
\end{figure}

Prior to matching, the workflow multiplies the elevation-point horizontal coordinates and elevation values by the US-survey-foot-to-meter factor. A regular \SI{25}{\meter} search grid then indexes the selected elevation points. For every unique network-geometry vertex \(v\) with horizontal coordinate \(x_v\), the assigned elevation is the unweighted mean
\begin{equation}
  \begin{aligned}
    z_v & = \frac{1}{|P_v|}\sum_{p\in P_v} z_p,                 \\
    P_v & = \{p: \lVert x_p-x_v\rVert_2 \leq \SI{50}{\meter}\}.
  \end{aligned}
  \label{eq:elevation}
\end{equation}
A vertex is supported when \(|P_v|\geq1\). The workflow does not substitute a distant nearest neighbor because such a fallback can yield plausible-looking but locally unrelated heights. A segment is retained only when every geometry vertex is supported. This rule preserves 99.24\pct of the source segments. The \SI{50}{\meter} radius trades spatial specificity for source coverage. Across all supported source vertices, the median number of contributing observations is 6, with a range of 1 to 80. Figure~\ref{fig:support-distance} shows that the median distance to the nearest contributing observation is \SI{10.14}{\meter}; the 90th and 99th percentiles are \SI{19.97}{\meter} and \SI{36.52}{\meter}, respectively. These diagnostics reduce but do not eliminate the risk of mixing vertically distinct observations, particularly near bridges, underpasses, and stairs.

\subsection{Segment Metrics}

For a retained segment with ordered vertices \((x_i,z_i)\), horizontal length is \(L=\sum_i\lVert x_{i+1}-x_i\rVert_2\). Positive and negative elevation changes are accumulated separately over all geometry steps:
\begin{equation}
  \begin{aligned}
    G^+ & = \sum_i \max(z_{i+1}-z_i,0), \\
    G^- & = \sum_i \max(z_i-z_{i+1},0).
  \end{aligned}
\end{equation}
These quantities are used to calculate directional costs. The dataset reports endpoint net change \(\Delta z=z_n-z_1\), endpoint-average grade \(100\Delta z/L\), and the maximum absolute local grade
\begin{equation}
  g_{\max}=100\max_{i:\,\ell_i\geq \SI{1}{\meter}}
  \frac{|z_{i+1}-z_i|}{\ell_i},
\end{equation}
where \(\ell_i=\lVert x_{i+1}-x_i\rVert_2\). If no step is at least \SI{1}{\meter} long, the reported maximum is set to zero. The one-meter cutoff is an arbitrary threshold that prevents extremely short horizontal steps from dominating this diagnostic field. It does not remove those steps from \(G^+\), \(G^-\), or routing cost, a distinction revisited in Section~\ref{sec:limitations}. Formulas below use grade ratios as decimals; exported grade fields are percentages.

\subsection{Routing Cost Functions}

The distance-only cost is \(C_0=L\). The other profiles use a transparent piecewise grade penalty. Let \(q^+=G^+/L\) and \(q^-=G^-/L\) be the cumulative ascent and descent ratios in a given direction. For threshold values \(g_0=0.05\) and \(g_1=1/12\), define
\begin{equation}
  \begin{aligned}
    h(q;w_m,w_s)={} & w_m\min\{\max(q-g_0,0),g_1-g_0\} \\
                    & +w_s\max(q-g_1,0).
  \end{aligned}
  \label{eq:penalty}
\end{equation}
The directional score is
\begin{equation}
  \begin{aligned}
    C=L\big[1&+h(q^+;w_{u,m},w_{u,s}) \\
         &+h(q^-;w_{d,m},w_{d,s})\big].
  \end{aligned}
  \label{eq:cost}
\end{equation}
The distance-only profile sets all four weights to zero. For moderate and steep grades, respectively, the comfort-oriented profile uses uphill weights of 8 and 30 and downhill weights of 2 and 8; the accessibility-sensitive profile uses uphill weights of 20 and 80 and downhill weights of 12 and 50. These asymmetric, dimensionless coefficients are selected scenario parameters based on the relative effort of climbing versus descending. The accessibility-sensitive profile increases both penalties, particularly above 8.33\pct. The 5\pct and 8.33\pct breakpoints correspond to recognizable values in ADA walking-surface and ramp specifications, but do not constitute a compliance assessment \citep{usaccessboard2010ada}. Because the costs use segment-aggregate \(G^+/L\) and \(G^-/L\), they do not directly optimize the maximum local grade reported in the released table. Accordingly, the profile is accessibility-sensitive rather than barrier-free: it minimizes cumulative slope-weighted cost but does not exclude a route solely because it contains, for example, a short, steep local step.

\subsection{Network and Route Computation}

To exemplify the workflow, the release includes a simple route-finding implementation of Dijkstra's shortest-path in Python. Each retained segment is stored once with endpoint identifiers \(a\) and \(b\); exact endpoint coordinates define routing nodes. For routing, the segment is expanded into two in-memory directed edges, with reversal exchanging \(G^+\) and \(G^-\) while preserving horizontal length and maximum absolute grade. The workflow assumes bidirectional traversal and preserves parallel segments by identifier. The algorithm computes a minimum-cost path for the selected directional cost. The O-D (origin-destination) pairs below are illustrative diagnostic cases whose coordinates are recorded in the accompanying analysis output.

\section{Results}

\subsection{Dataset Characteristics}

The analytical release contains \num{187657} endpoint nodes and \num{313184} physical segments, equivalent to \num{626368} directed edges. The segment-length median is \SI{32.07}{\meter}; the 95th percentile is \SI{218.96}{\meter}. Node elevation ranges from \SI{-0.33}{\meter} to \SI{121.09}{\meter}, with a median of \SI{12.97}{\meter}. Median absolute endpoint-average grade is 0.58\pct, and the 95th percentile is 4.74\pct. The maximum local absolute grade has a median of 0.88\pct and a 95th percentile of 10.60\pct. In total, \num{40064} segments (12.79\pct) contain a reported local grade of at least 5\pct, and \num{21430} (6.84\pct) reach at least 8.33\pct. By contrast, \num{14895} segments (4.76\pct) receive a slope penalty in at least one direction. The lower count is not contradictory since the local maximum summarizes individual geometry steps, whereas the cost threshold is applied to segment-aggregate \(G^+/L\) or \(G^-/L\).

\begin{figure}[t]
  \centering
  \includegraphics[width=0.90\linewidth]{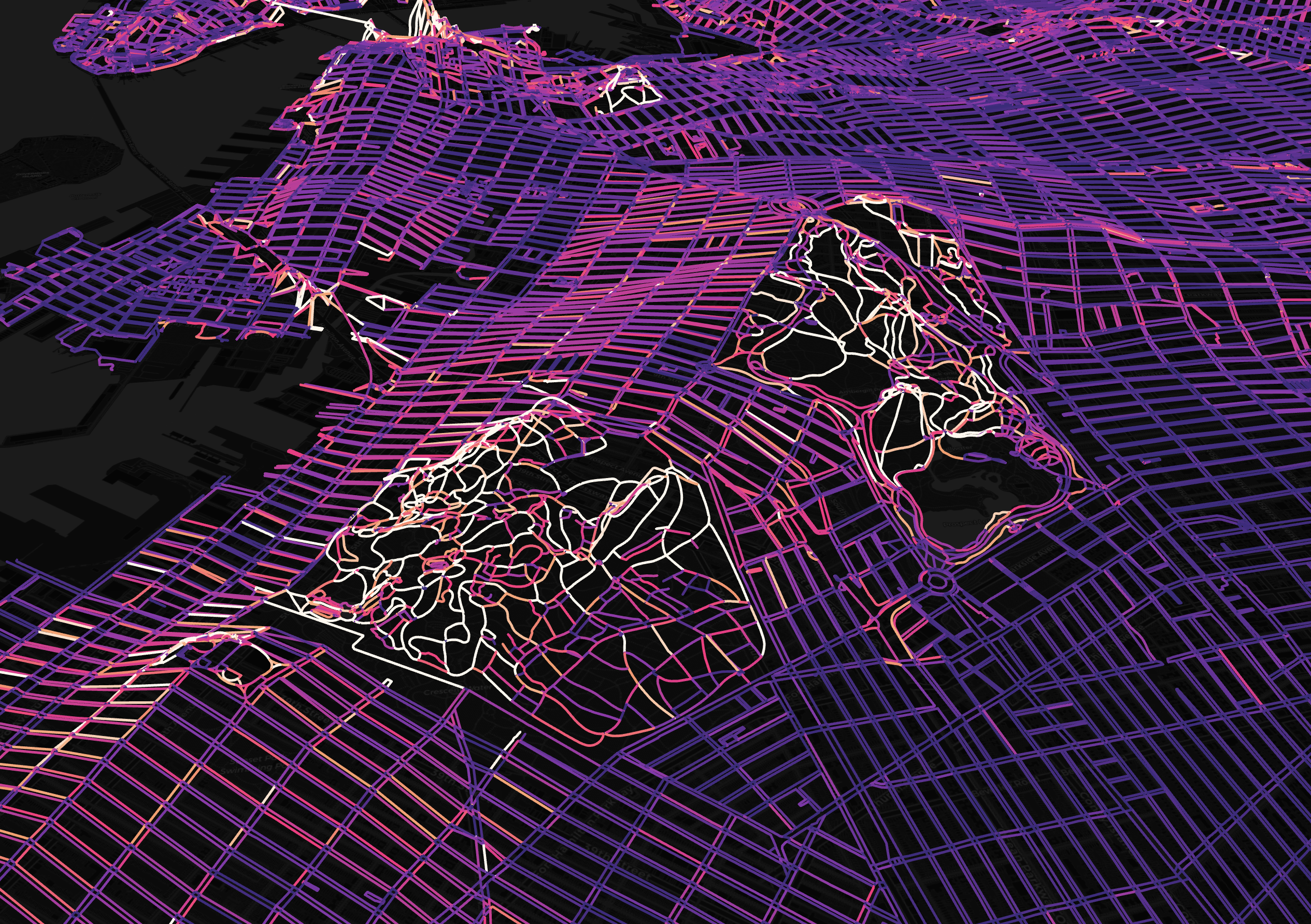}
  \caption{Vertically exaggerated view of the \dataset network looking northwest from Green-Wood Cemetery and Prospect Park. Color encodes maximum reported absolute local grade, with brighter tones indicating steeper values. Park and open-space paths visibly follow natural terrain. Park Slope also shows greater grade variation than Crown Heights and Prospect Lefferts Gardens, illustrating that local grade can vary substantially across adjoining neighborhoods.}
  \label{fig:elevation-slope-view}
\end{figure}

\subsection{Consistency Checks}

Repeating the analysis on the archived source data reproduced the released coverage. Including all records in the Planimetric \texttt{ELEVATION} class instead skewed the resulting elevations near buildings, confirming the need to filter elevation feature codes before matching. Review identifies \num{470} segments (0.15\pct) with a \emph{reported} local absolute grade at or above 100\pct. At Dyckman Fields, one \SI{0.34}{\meter} geometry step changes \SI{4.74}{\meter} in elevation. Because no step on that segment passes the one-meter reporting threshold, its maximum local grade is reported as zero, while its accumulated gain still produces an extreme cost. Such records may represent vertical connections, but they may also result from geometry artifacts or mismatched elevation support. The current filtered maximum-grade field therefore does not expose every extreme value that can influence cost, and such cases require explicit review and handling before real-world use.

\subsection{Route Comparison Case Studies}

\begin{figure}[t]
  \centering
  \includegraphics[width=\linewidth]{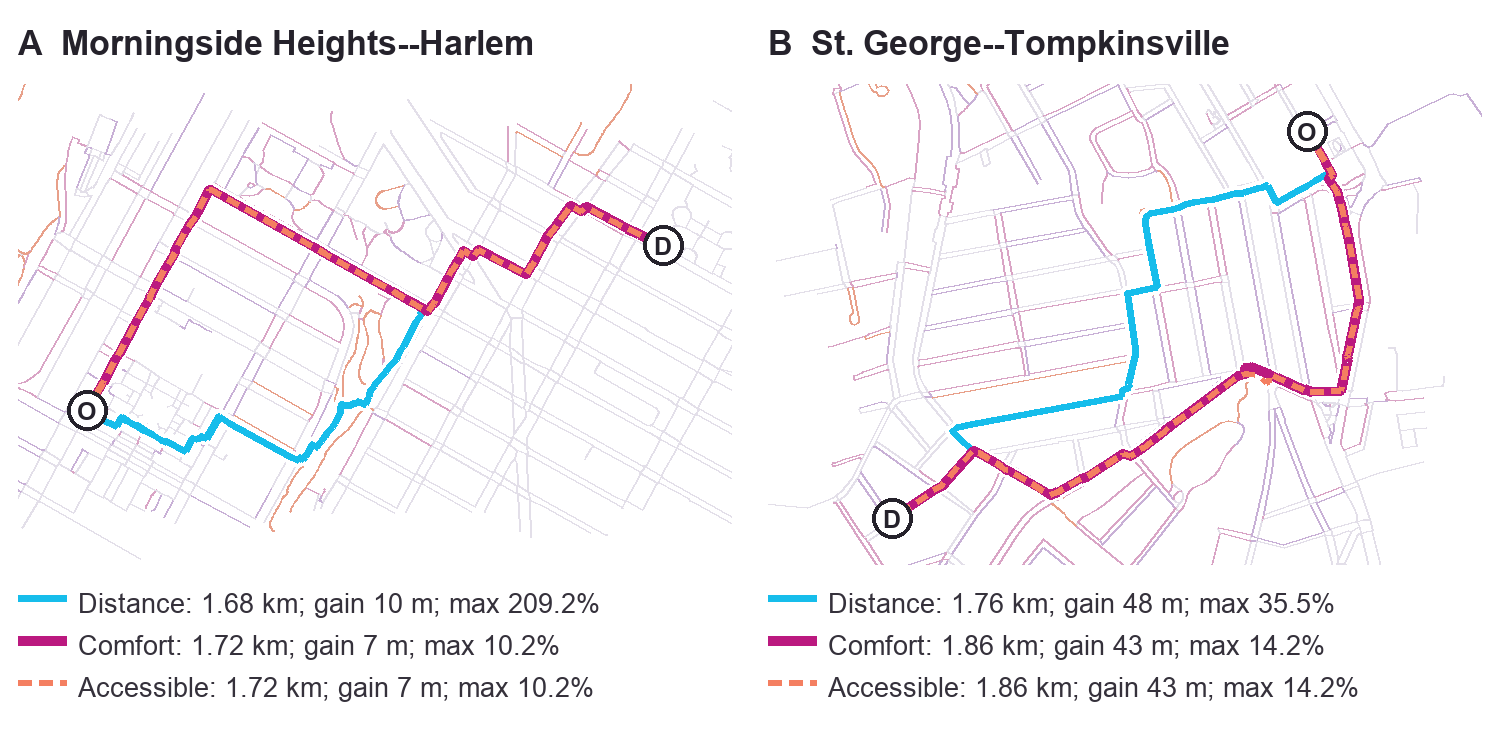}
  \caption{Illustrative route comparisons reconstructed from the released graph. Panel A shows a diagnostic case in Morningside Heights/Harlem, where an implausibly steep elevation-assigned edge changes route choice. Panel B shows a more plausible distance-grade trade-off in St.~George/Tompkinsville. Gain labels reproduce the current application's endpoint-summed metric, whereas routing costs use changes across all intermediate geometry vertices.}
  \label{fig:routes}
\end{figure}

Figure~\ref{fig:routes} compares the three profiles for two O-D pairs. In Staten Island, the distance-minimizing route is \SI{1.76}{\kilo\meter}. The routing output reports \SI{48.49}{\meter} of endpoint-summed gain and a maximum local grade of 35.5\pct. Recalculation across all intermediate geometry vertices gives \SI{61.48}{\meter} of gain, which, together with full-geometry loss, enters the assigned cost. The comfort-oriented route is 5.5\pct longer but reduces full-geometry gain to \SI{43.30}{\meter} and the maximum reported local grade to 14.2\pct. The accessibility-sensitive route follows nearly the same corridor.
\\
The Morningside Heights distance route includes an edge with a reported grade of 209.2\pct, an implausible value for a typical walkway. Both slope-sensitive profiles avoid this edge by choosing a route that is 2.3\pct longer, reducing the maximum reported grade to 10.2\pct. Full-geometry gain falls from \SI{19.59}{\meter} to \SI{7.00}{\meter}. The selected paths therefore minimize the assigned costs as specified, although the underlying elevation-derived grade may be inaccurate.
\\
Extreme values can arise when a \SI{50}{\meter} window mixes ground and bridge-deck observations, when short features amplify noise, or where stairs and stacked connections represent valid vertical facilities. The cases demonstrate both the dataset's diagnostic value and the need for edge review before accessibility routing.

\section{Discussion}

\subsection{Limitations}
\label{sec:limitations}

Assigned vertex elevations are unweighted local means of municipal spot observations rather than accurate survey-grade measurements. The \SI{50}{\meter} window was selected to address the sparsity of source observations, but it can smooth terrain, combine observations from vertically distinct features, and propagate erroneous source values. The workflow converts the EPSG:2263 horizontal coordinates to meters by unit scaling rather than explicitly transforming them to EPSG:6538; any residual offset between NAD83 and NAD83(2011) has not been quantified.
\\
The complete-support criterion can fragment the network where elevation coverage is sparse. The released graph has 256 connected components. Its largest contains \num{158110} nodes (84.25\pct) and \num{266698} physical edges; the next two largest correspond to Staten Island and the Rockaway Peninsula, which are geographically separated from the largest component. At the Marine Parkway and Cross Bay bridges, five source vertices have their nearest selected elevation observations 1.27-5.49 meters beyond the \SI{50}{\meter} radius. Supporting all five would require a radius of approximately \SI{55.5}{\meter}, but a larger window would also increase the risk of mixing vertically distinct observations and was not adopted for the release.
\\
The workflow assumes that the source graph represents valid, bidirectional walkable connectivity. Elevation enrichment cannot establish whether a link is legally accessible, temporarily obstructed, or correctly connected across stacked infrastructure. The released cost logic also does not distinguish streets, ramps, stairs, and bridges. The graph contains two self-loops and \num{4363} endpoint pairs joined by parallel segments.
\\
The grade-sensitive routing profiles are heuristic and sensitive to the segmentation of source features. Because the penalty is nonlinear, splitting or merging the same physical path can change the resulting cost, and a steep local step can be diluted within a long segment. Finally, no systematic comparison with survey-grade sidewalk profiles or field observations has established error rates for assigned elevation or grade.

\subsection{Future Research}

Future work should evaluate more selective geographic and topological matching, test sensitivity to the search radius, and compare assigned elevations with independent survey or field observations. The release can also support tests of how terrain changes pedestrian catchments, centrality, and foot-traffic models. Empirically calibrated or mobility-specific cost functions could then replace the illustrative scenarios \citep{paez2020comparing,hosseini2024inclusive}.

\section{Conclusion}

\dataset makes terrain an explicit, inspectable part of a citywide pedestrian graph for New York City. It documents how local elevations are assigned, how direction-specific grade scores are constructed, and how route choice changes across the three routing profiles. The release also exposes unresolved cases where topography, vertical infrastructure, and data artifacts are hard to distinguish. Its immediate value therefore lies in enabling reproducible terrain-aware analysis and providing a concrete basis for improving how pedestrian networks represent elevation.

\section*{Data and code availability}

The data, code, QGIS project, and web application are available at \url{https://github.com/RELNO/gridnberg}. The public application is deployed at \url{https://arielnoyman.com/gridnberg/}. The analytical tables are \path{outputs/nyc_ped_net_radius50m_routing_network.csv} and \path{outputs/nyc_ped_net_radius50m_routing_nodes.csv}; the companion GeoPackage preserves the intermediate segment geometry needed to inspect vertex-dependent metrics. The source NYCWalks network and NYC Planimetric elevation points remain governed by their respective providers and are cited separately.

\section*{Acknowledgments}

The author thanks Andres Sevtsuk and collaborators for their pedestrian-network research and datasets on which \dataset builds, and the City of New York for maintaining the Planimetric Database. The author also thanks J.T., who earned the right to observe sidewalk slopes before the rest of us.

\balance
\printbibliography

\end{document}